# Webcam Eye Tracking: Study Conduction and Acceptance of Remote Tests with Gaze Analysis

Investigation of Data Quality and Tester Satisfaction in a Gaze Cueing Study

Sezen Lim, Tina Walber, Christoph Schaefer, Lena Riehl

Eyevido GmbH, info@eyevido.de

Webcam eye tracking for the collection of gaze data in the context of user studies is convenient – it can be used in remote tests where participants do not need special hardware. The approach has strong limitations, especially regarding the motion-free nature of the test persons during data recording and the quality of the gaze data obtained. Our study with 52 participants shows that usable eye tracking data can be obtained with commercially available webcams in a remote setting. However, a high drop off rate must be considered, which is why we recommend a high over-recruitment of 150%. We also show that the acceptance of the approach by the study participants is high despite the given limitations.

## 1 INTRODUCTION

In the wake of the global covid-19 pandemic, user labs temporarily had to close for health and safety reasons. As a result, the relevance of remote testing took on a new significance in the field of usability research. Conventional infrared eye tracking systems provide high-quality data but are only accessible in user labs. In addition to being cost-intensive, organizing on-site appointments with testers can be time-consuming.

Webcam eye tracking is a promising alternative to overcome these financial and organizational difficulties. This method does not require any special technical equipment or time constraints. Test users only need a standard webcam. The data recording can take place in the web browser. This opens up the opportunity of reaching a much higher number of participants as there is no need to visit the lab and attend the test one after the other. However, webcam studies often experience challenges with accuracy and feasibility, which may limit their overall usefulness.

Our study demonstrates the feasibility of remote webcam eye tracking studies by investigating the gaze cueing effect, a well-known eye tracking phenomenon. Data was collected from 52 participants split into two groups. Our analysis shows that the eye tracking data obtained by webcams is accurate enough to demonstrate the gaze cueing effect. In addition to analyzing the obtained gaze data, we investigated how many participants dropped out before completing the study and collected user feedback in a questionnaire.

*The contribution of this work is: 1) we show that known effects in human attention distribution can be measured with webcam eye tracking, 2) we discuss the recruitment process carried out, and 3) show that the acceptance of webcam eye tracking is high among the remote participants.*

## 2 RELATED WORK

Existing research shows mixed findings regarding the performance of webcam eye tracking. Consistently cited insufficient factors include reduced accuracy compared to eye tracking data collected in user labs and the influences of a test environment outside the lab [1]. Such problems can stem from the quality of the webcam used, the lighting conditions, as well as the posture of the participants when no study supervisor is present [2].

Despite these limitations, many authors consider webcam eye tracking a promising possibility for the future. Whether webcam eye tracking is an applicable method depends on the purpose of the study or the design itself [1] [2] [3] [4]. In a study comparing data quality for larger images centered on the screen [1], webcam eye tracking was only inferior to the infrared version when smaller images were in the viewer's periphery. A study on web design also argues for a comparability of the two systems [4]. Examinations with the webcam resulted in similar implications for the usability of a website design as a test with infrared systems. Cognitive research also provides evidence here, depending on the surrounding conditions. [2] compared three different tasks (fixation, free viewing, and tracking) under user lab conditions and remote. A JavaScript algorithm [3] [5] was used together with the webcams of the participants. The evaluation showed increased variance in the online data and longer sessions, but still speaks of a comparable accuracy. Thus, whether the use of remote webcam eye tracking should be deemed useful depends not only on the technical means alone, but also on the deployment and design of the research in question, resulting in new implications for further exploration.

### 3 STRUCTURE OF THE WEBCAM STUDY

#### 3.1 Subjects and Eye Tracking System

The study subjects were recruited by a newsletter, classified ads, and social media. Following this, the subjects registered on a booking portal and selected a day on which they could participate, since this was predefined by the booking portal. After the study director confirmed the registration, the subjects received a participation invite link via e-mail.

The study was conducted with EYEVIDO Lab, a cloud software for web-based eye tracking analyses, by Eyevido GmbH. Subjects participated in the web tool on their PC using their own webcam. The software enables the creation and analysis of screen-based eye tracking studies. It offers different questionnaire elements for the design, and different analysis tools like AOIs, mouse-data and gaze plots. The tester software records the screen content and the eye movements of the participants, stores and transmits all data in the cloud.

#### 3.2 Stimuli and Task: Gaze Cueing Effect

Humans and even primates follow the line of sight of their conspecific [6]. This is shown, for example, when people in pictures look in a certain direction and viewers of the image follow the direction of view before turning to the rest of the image. This principle also works with other visual cues such as arrows [7] [8] [9]. Our goal was to generate the known effect in the subject's gaze behavior and record it via webcam eye tracking so that it becomes visible in the data in order to provide evidence of the usefulness of webcam eye tracking.

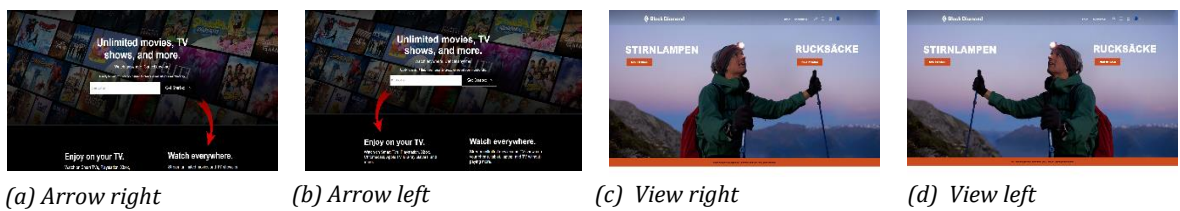

*(a) Arrow right*   *(b) Arrow left*   *(c) View right*   *(d) View left*

*Figure 1: Examples of screenshots (website: netflix.com, taken on 14.04.2022) featuring arrows as cue stimuli pointing towards right(a) and left(b) content blocks. And screenshots (website: blackdiamondequipment.com, taken on 07.04.2022) featuring person as a cue.*

The stimuli used were screenshots of five different websites. Three web pages contained a cue stimulus in the form of a depicted person looking either to the left or right, and two web pages had an arrow pointing to specific web page elements (Figure 1). The screenshots were manipulated by means of image processing in such a way that the left and right screenshots contained balanced content blocks (information content, size of text, colors, and images) that were positioned almost symmetrically to avoid interfering attention processes. The study was conducted in two groups (A and B). The content and course of the study were identical for both groups, except that the cues of groups A and B were positioned in exactly opposite directions.

#### 3.3 Implementation

Each participant received a link to a PDF with detailed instructions. Subjects were instructed not to move, position light sources and their webcam front and center, and not to wear glasses. Via a link in the PDF, the participants were forwarded to a web form to accept the terms and conditions and privacy policy. After accepting the terms, the subjects were forwarded to the study where they had to input their age and gender. After that, a 9-point calibration process began. Subjects had to achieve a predefined high calibration quality to qualify for a participation in the webcam eye tracking study. How often subjects had to calibrate to reach the quality was not specified.

Before the stimuli were presented, test users were instructed to "figure out what is offered on the site". After eight seconds the viewing was automatically stopped. This short time frame was chosen because, attentional effects typically appear after a maximum of 300ms in standardized reaction time experiments [10]. A dropdown question automatically followed in which subject had to select from five alternatives regarding the type of content that was previously presented on the website. After submitting their answer, they were automatically transitioned to the following screenshot. This study design served to motivate the test users. During the study, testers were repeatedly given written instructions not to move. After presenting the screenshots and the corresponding questions, subjects were asked general questions about the implementation of the study. Lastly, subjects provided information on how they would like to receive the compensation of 5 € for their efforts.

## 4  RESULTS

After the recruitment process, 52 subjects were admitted to the study. Of these, 61 % were female and 39 % male. The mean age of subjects was 27.8 years with a SD of 6.7. Comprehension questions regarding the content of the screenshots were answered 98.8 % correct, thus it can be concluded that subjects were sufficiently attentive and concentrated.

### 4.1  Evaluation of Webcam Eye Tracking Data

For the purposes of evaluating the results, apparent invalid eye tracking data sets of subjects were excluded. An incorrect data set occurs when the webcam is not able to capture the correct eye position and gaze direction of the tester. This can be caused by the subject's head movements, but can also result from external influences, such as unfavorable lighting conditions. Fourteen clearly invalid datasets (38.9% of all datasets) were identified and excluded from further analysis.

To analyze the gaze on the screenshots, two areas of interest (AOIs) were created on each stimulus, one for each information block cued in group A resp. B. If the cue stimulus is pointing at a content block, it is labeled as $AOI^{CUED}$, the other block is labeled as $AOI^{CTRL}$ (*Figure 2 a, b*). The total number of fixations of all participants on those areas are calculated. Subsequently, the percental distribution of fixations between the two AOIs was determined for each stimulus.

On screenshots showing a left-pointing cue stimulus, all values for the left and all values for the right AOI were summed, and the mean value was determined for each side individually *(Figure 2 c)*. The same procedure was repeated for screenshots with a right-pointing cue stimulus *(Figure 2 d)*. $AOI^{CUED}$ achieved a greater number of fixations compared to $AOI^{CTR}$, regardless of whether the corresponding content blocks were positioned on the left or right. Overall, there is a higher percentage in the left AOIs. This can be attributed to the typical left-to-right reading pattern. Thus, the directions of the testers gaze could be measurably influenced by the cue stimulus. Webcam eye tracking can thereby evidentially generate meaningful eye tracking data, which can be used for the evaluation of eye tracking studies.

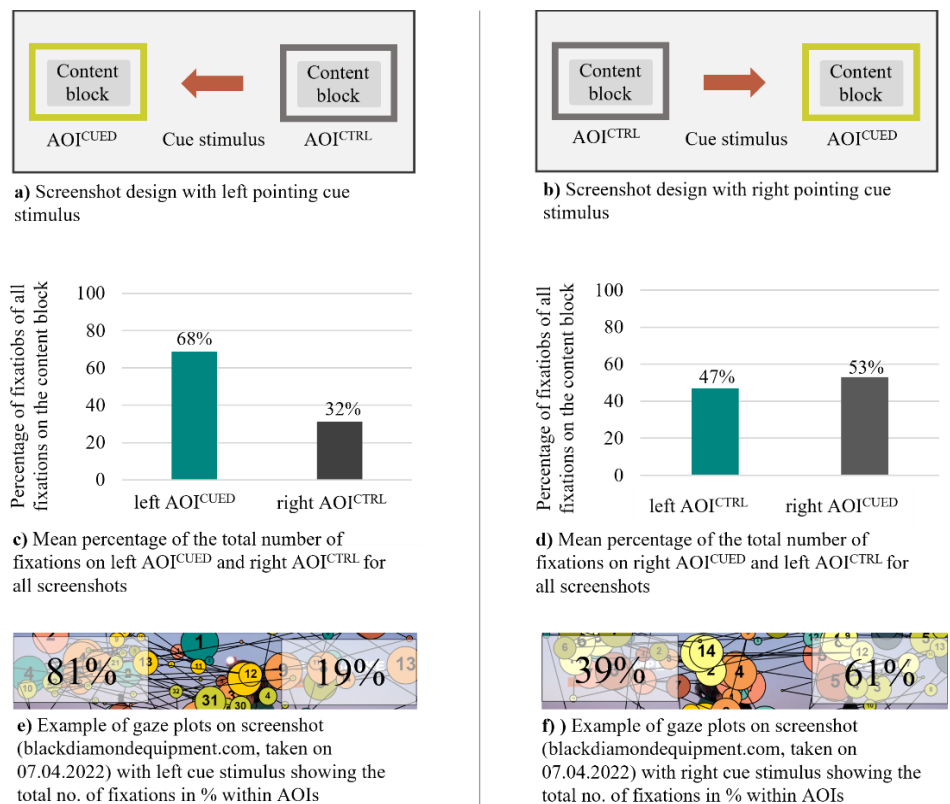

*Figure 2: Evaluation scheme for gaze cueing stimuli.*

## 4.2 Learnings from the Recruitment Process

Remote testers were recruited via a newsletter, classified ads, and social media *(Figure 3)*. A total of 57 remote testers signed up on a booking page to participate in the online webcam study. Subscribers to the newsletter received an email a few days prior to the start of the webcam eye tracking test with a link to the booking page to sign up for the test. 86 eligible testers were informed about the upcoming test by the newsletter, of whom 47 opened the email, and 21 clicked on the link. 36.8 % of testers were recruited through the newsletter. 63.2 % of the remote testers were recruited via classified ads and social media. Classified ads were viewed a total of 191 times and the social media group had 4,735 members when the call for testers was posted into it. It is unclear how many remote testers signed up from each platform.

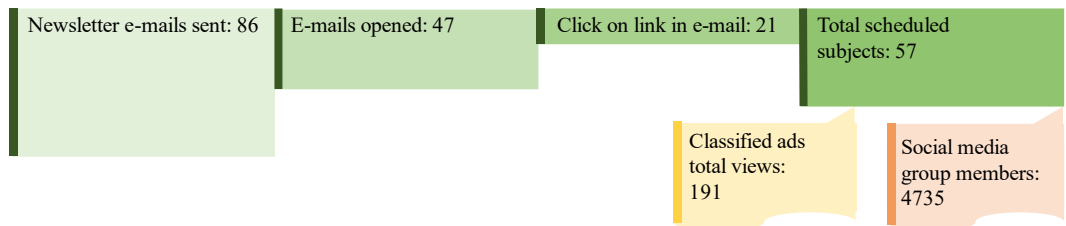

*Figure 3 Subjects recruited from newsletter, classified ads, and social media.*

A possible limitation in the registration process was the booking of appointments to participate in the test. After confirming their registration on the booking page, testers could only participate on the scheduled day and were not able to select an exact time slot. This could have caused general uncertainty and thus led to drop off cases. Unfortunately, it is not possible to retrospectively track how many testers dropped off while signing up for the test. From the recruitment process up to the evaluation of the study, it can be concluded that a large number of potential testers must be approached in order to obtain sufficient participants. Depending on the platforms and media used to target testers, the total number of recruited testers should be at least four times greater than the number of desired participants to arrive at a sufficient sample size.

## 4.3 Survey of Remote Webcam Eye Tracking Studies

Out of 57 test registrations, 54 individuals accepted the terms and conditions and privacy policy, and 52 testers subsequently participated in the webcam test. 30.8 % of the participants could not pass the calibration because the eye tracking was too inaccurate. Among the remaining participants, 38.9 % hat invalid eye tracking data sets. Consequently, of the 57 recruited testers, only 22 valid data sets could be obtained *(Figure 4)*. Ultimately, the data of eight males and 14 females were included in the analysis.

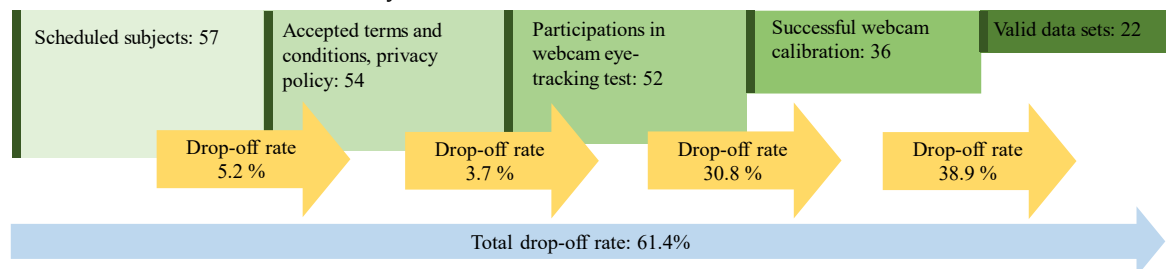

*Figure 4: Number of subjects per recruitment step.*

The total drop rate from signing up to participate in the test to the obtaining webcam eye tracking data is 61.4 %. This high proportion of non-existent or invalid data sets must be considered in webcam eye tracking studies and compensated by an over-recruitment of about 150 %.

## 4.4 Acceptance of Webcam Eye Tracking by Subjects

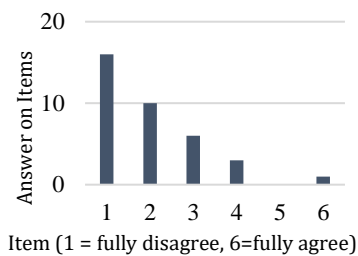

Figure 5: "I found the eye tracking bothersome"

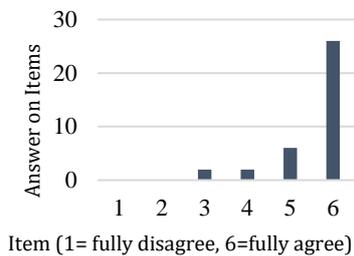

Figure 6: "I can imagine participating in a remote webcam eye tracking study in the future"

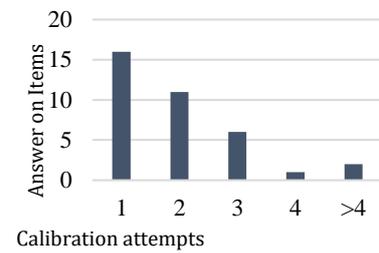

Figure 7: "How many calibration attempts did you need?"

To investigate how the participants experienced the webcam eye tracking, they answered different questions on a six-point Likert-Scale (1 = fully disagree, 6 = fully agree). To examine the process of calibration, which was needed for reliable data quality, subjects were asked about the amount of calibration attempts, till they succeeded. 75 % of subjects needed less than two attempts *(Figure 7)*. This mirrors the 75 % of subjects, who rated the calibration process as uncomplicated in a separate question. The data reflects the subjects hardly felt disturbed during eye tracking recordings: 89 % did not agree with the statement "I found the eye tracking *bothersome*" *(Items 1-3, Figure 5).* 95 % even stated that they would like to participate in a remote webcam eye tracking study again in the future *(Items 4- 6, Figure 6).*

Additionally, subjects were asked to provide their written feedback. The responses were categorized by valence. Five subjects addressed deficits and therefore these statements were classified as negative. For example, eye tracking was described as "very difficult". Seven were classified as neutral. This included feedback that was not related to the feasibility of webcam eye tracking or that influenced the quality of the data. Three positive statements like: "fast" or "great" could be identified. Even if more negative feedback was written in the free text form than positive, this is not to be evaluated negatively, since deficits are more likely to be addressed than intuitive processes.

It can be stated that the calibration process was mostly perceived as uncomplicated, and that webcam eye tracking was not perceived as irritating. Consequently, there is no need to provide high compensation to the subjects to ensure their motivation. The relatively low compensation amount of 5 € seems to have been sufficient.

## 5 SUMMARY AND OUTLOOK

This study shows that remote webcam eye tracking delivers valid data and insights into user behavior if a high failure rate due to quality problems is considered. The acceptance of webcam eye tracking by the participants is high, despite the demands placed on them.

In our future work, we will investigate how subjects can be better supported to create favorable external conditions for webcam eye tracking. The idea is to offer a prior data quality check to verify the light conditions. Furthermore, it should be monitored in real time whether the head in front of the screen is held constant. Immediate visual cues should remind the subject not to move. Regarding the analysis of the obtained data, it will be investigated whether the analysis procedures must be adjusted to the lower quality of the eye tracking data. Gaze data should no longer be analyzed on the stimulus but relatively per gaze path to measure a general orientation of attention even when the data is scaled or shifted.